\documentclass[9pt,twocolumn,twoside]{osajnl}
\usepackage{siunitx}
\usepackage[percent]{overpic}

\usepackage[overlay,absolute]{textpos}
\newcommand\PlaceText[3]{%
	\begin{textblock*}{10in}(#1,#2)
		#3
	\end{textblock*}
}%

\journal{ol} 

\setboolean{shortarticle}{true} 

\title{Emission beyond 4~\si{\um} and mid-infrared lasing in a dysprosium-doped indium fluoride (InF$_3$) fiber}

\author[1,*]{Matthew R. Majewski}
\author[1]{Robert I. Woodward}
\author[2]{Jean-Yves Carre\'e}
\author[2]{Samuel Poulain}
\author[2]{Marcel Poulain}
\author[1]{Stuart D. Jackson}

\affil[1]{MQ Photonics, School of Engineering, Macquarie University, North Ryde, NSW 2109, Australia}
\affil[2]{Le Verre Fluor\'e, Campus KerLann, F-35170 Bruz, Brittany, France}

\affil[*]{Corresponding author: matthew.majewski@mq.edu.au}

\ociscodes{(140.3510) Lasers, fiber (140.3070) Infrared and far-infrared lasers}

\doi{\url{http://dx.doi.org/10.1364/OL.43.001926}}

\begin{abstract}
Optical emission from rare-earth-doped fluoride fibers has thus far been limited to less than 4~\si{\um}. 
We extend emission beyond this limit by employing an indium
fluoride (InF$_3$) glass fiber as the host, which exhibits an
increased infrared transparency over commonly used
zirconium fluoride (ZBLAN).
Near-infrared pumping of a dysprosium-doped InF$_3$ fiber results in broad emission centered around 4.3~\si{\um}, representing the longest emission yet achieved from a fluoride fiber. 
The first laser emission in an InF$_3$ fiber is also demonstrated from the 3~\si{\um} dysprosium transition. 
Finally, a frequency domain excited state lifetime measurement comparison between fluoride hosts suggests that multiphonon effects are significantly reduced in indium fluoride fiber, paving the way to more efficient, longer wavelength lasers compared to ZBLAN fibers.  
\end{abstract}

\setboolean{displaycopyright}{true}

\begin{document}

\maketitle

\PlaceText{25mm}{9mm}{Vol. 43, Issue 8, pp. 1926--1929 (2018); https://doi.org/10.1364/OL.43.001926}

Sources of laser emission in the mid-infrared (MIR) spectral region are finding a growing number of advanced applications across a variety of fields. 
The 3--5~\si{\um} span in particular has attracted significant interest, as it covers several key ro-vibrational absorption features of molecules containing CH, and CO, offering distinct potential for spectroscopic and materials processing applications.
Additionally, this window contains regions of exceptionally high atmospheric transmission, allowing for sources here to be of potential use in light detection and ranging (LIDAR) or infrared countermeasures for example.

To meet the laser source needs of these applications, rare-earth-doped fiber lasers have emerged as promising candidates. 
Fiber-based systems offer high-brightness in a compact package, and with proper design, the potential for efficient generation of high output power.
As mid-infrared transitions of rare earth ions are strongly quenched by multi-phonon relaxation in conventional silicate glasses due to their high maximum phonon energy ($\sim$1100 cm$^{-1}$), the development of fiber laser sources in the MIR has required the accompanying development of alternative soft glass hosts~\cite{jackson2012towards}.
The most successful of these have been zirconium fluoride based glasses (fluorozirconates)~\cite{poulain1975verres} principally comprising ZrF$_4$ as the glass former, in addition to network modifiers and stabilizers such as BaF$_2$ and NaF; the most common composition is ZrF$_4$--BaF$_2$--LaF$_3$--AlF$_3$--NaF (ZBLAN) with $\sim$600~cm$^{-1}$ phonon energy~\cite{zhu2010high}.  
ZBLAN fiber lasers doped with dysprosium have shown emission spanning 2.8 to 3.4~\si{\um}~\cite{Majewski:18}, while an erbium-doped system achieved lasing up to 3.78~\si{um}~\cite{henderson2016versatile}.
The longest wavelength from a ZBLAN fiber laser to date was based on holmium doping, with emission at 3.9~\si{\um}, albeit with cryogenic cooling as a necessity~\cite{schneide1997characterization}. 
Further extension of emission wavelength in a zirconium fluoride glass has not been possible due to the increased influence of multiphonon relaxation of excited states for longer wavelength transitions (i.e. quenching), in addition to exponentially increasing loss beyond $\sim$3.8~\si{\um}. 
Continued development of fiber laser sources covering the entire 3--5~\si{\um} window thus requires the focus to shift from ZBLAN to glasses with further reduced phonon energy. 

Chalcogenide-based glass fibers with phonon energies as low as $\sim$300~cm$^{-1}$ have shown promising ongoing development~\cite{sojka2017mid}, though to date, chalcogenide fiber laser action in the MIR has proven elusive, attributed to impurity multiphonon relaxation, high background loss and the inability to support high rare-earth doping concentrations.
Therefore, it is preferable to explore other fluoride-based fibers, leveraging the benefits of maturing fluoride glass/fiber fabrication and processing technology following the success of ZBLAN.

Fluoride glasses with indium fluoride (InF$_3$) as the primary glass former (fluoroindates) are currently emerging as promising MIR materials with a lower phonon energy than ZBLAN of $\sim$509~cm$^{-1}$~\cite{almeida1993vibrational}, broader transparency window and the ability to be drawn into low-loss fiber. 
Indium fluoride already has enabled increased spectral coverage of fiber laser sources as a nonlinear medium through soliton self-frequency shift~\cite{tang2016generation}, and supercontinuum~\cite{gauthier2016mid} generation phenomena.  
However, these nonlinear effects are based on passive InF$_3$ fiber, and while lasing on the 3.9~\si{\um} transition of Ho$^{3+}$ has been demonstrated recently in bulk InF$_3$~\cite{berrou2015mid}, neither MIR emission nor lasing has yet been shown in active InF$_3$ fiber.
In this letter we demonstrate for the first time to our knowledge, both MIR lasing and emission beyond 4~\si{\um} from a rare-earth-doped InF$_3$ fiber. 

\begin{figure}[!htbp]
	\centering
	\includegraphics[scale=0.5]{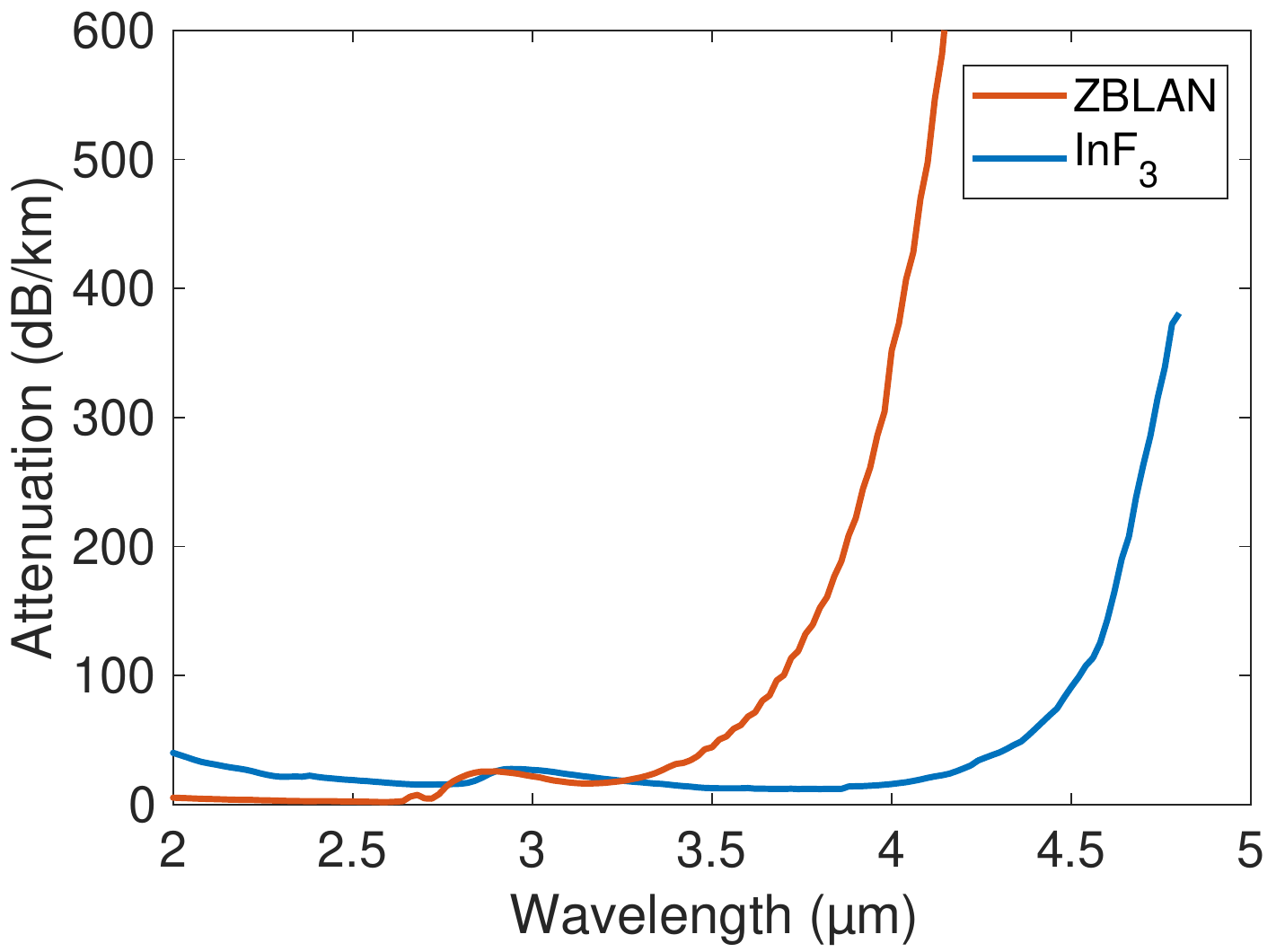}
	\caption{Measured attenuation of passive fluoride fibers.}
\label{fig:fiber_attn}
\end{figure}

Indium fluoride fiber used in this work is fabricated by Le Verre Fluor\'e and is synthesized from high purity fluorides with very low concentration of transition metals. 
The chemical composition is based on the IZSB glass (40~InF$_3$--20~ZnF$_2$--20~SrF$_2$--20~BaF$_2$)~\cite{messaddeq1991stabilizing,maze1996fluorinated} with additives to adjust refractive index, drawing temperature and thermal expansion.
The batch of microcrystalline powders is thoroughly mixed and melted in a very dry atmosphere, then the preform is made using a rod-in-tube process that ensures constant diameters and core to cladding ratio. 
Core diameter is determined from the values of numerical aperture and cut-off wavelength, which in this work is 0.16 and 2.6~\si{\um} respectively, resulting in a core diameter of 12~\si{\um}.
Fiber drawing is implemented under an inert atmosphere with a UV curable epoxyacrylate coating.
This process results in a fiber which exhibits substantially reduced attenuation beyond 4~\si{\um} as compared to a ZBLAN fiber of similar geometry as shown in Fig.~\ref{fig:fiber_attn}.

\begin{figure}[!htbp]
	\centering
	\begin{overpic}[scale=0.45]{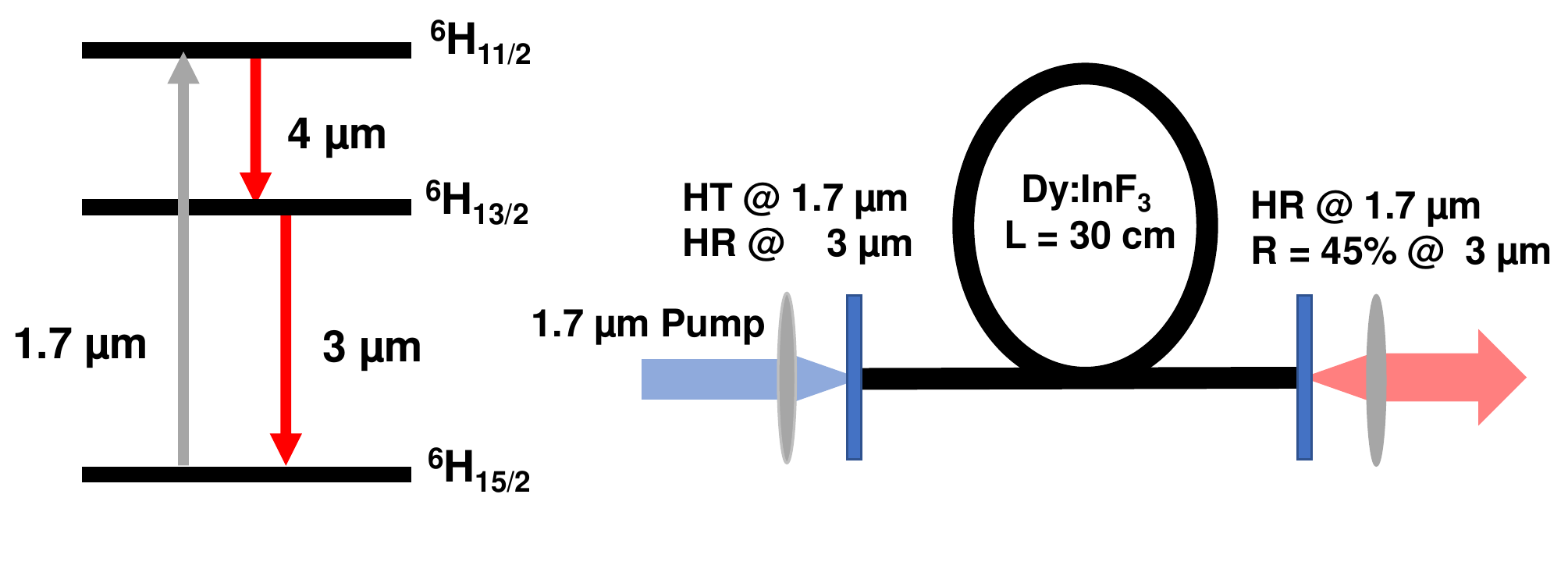}
		\put(14,0){\textbf{(a)}}
		\put(66,0){\textbf{(b)}}
	\end{overpic}

	\caption{(a) Simplified energy level diagram of dysprosium showing ground state pump absorption and both mid-IR radiative transitions. (b) Experimental schematic (the cavity mirrors are removed for the 4~\si{\um} emission measurement). }
	\label{fig:level_setup}
\end{figure}

To explore the potential of InF$_3$ fiber for mid-infrared laser development, we consider dysprosium as the active ion.
The motivation for this choice can be understood by inspection of the simplified energy level diagram in Fig.~\ref{fig:level_setup}(a).
Dysprosium exhibits two MIR transitions in the 3--5~\si{\um} window which are close to, and in the case of the 3~\si{\um} transition, explicitly terminated on, the ground state. 
This implies the potential for low quantum defect systems, with corollary high efficiency, as evidenced by the recent demonstration of a 3~\si{\um} fiber laser with up to 73\% efficiency~\cite{Woodward:18}.
Furthermore, the $^6H_{11/2}$ upper state can be pumped directly by a 1.7~\si{\um} source, which we have previously constructed for the recent demonstration of broadly tunable lasing from the $^6H_{13/2}\rightarrow^6H_{15/2}$ transition in Dy:ZBLAN~\cite{Majewski:18}.
In brief summary, this pump source consists of a Raman fiber laser pumped by the nominally 1570~nm output of a diode-pumped Er/Yb double clad fiber laser.

The lifetime of the lower state of the 4~\si{\um} transition ($^6H_{13/2}$) is longer than the upper state ($^6H_{11/2}$)~\cite{gomes2010energy}, which indicates that this transition should be of a self-terminating nature.
As a possible route to achieving 4~\si{\um} continuous-wave (CW) laser emission from dysprosium in indium fluoride, cascade lasing on the 3~\si{\um} transition has been proposed~\cite{quimby2013dy}. 
Therefore we first endeavor to demonstrate efficient lasing on the ground state terminated $^6H_{13/2}\rightarrow^6H_{15/2}$ transition.
We achieve this using the 1.7~\si{\um} pump source and the simple linear cavity arrangement depicted in Fig.~\ref{fig:level_setup}(b).
Pump light is injected into the core of a 30~cm length of singly clad step-index Dy(2000ppm):InF$_3$ fiber with an aspheric lens, and the cavity is closed by butt-coupled dichroic mirrors on both ends: the input mirror is highly reflective and the output coupler is 45\% reflective for $\sim$3~\si{\um} light.
The fiber length is chosen such that nominally 70\% of the injected pump light is absorbed in a single pass, while the output dichroic reflects all unabsorbed pump.
This results in near total pump absorption, and high inversion per unit length, facilitating a minimal value of oscillation threshold.

\begin{figure}[!htbp]
	\centering
	\begin{overpic}[scale=0.5]{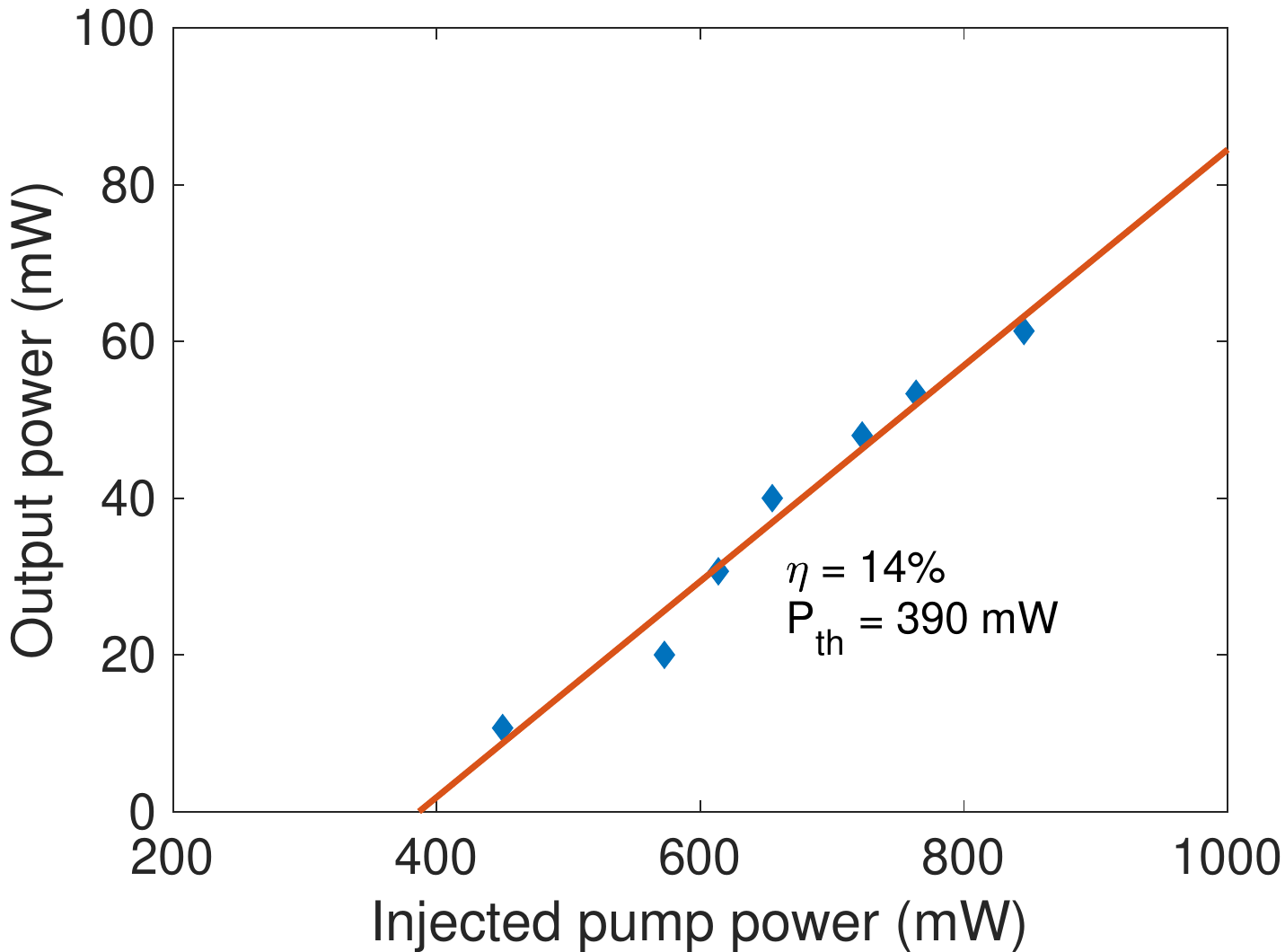}
		
		\put(17,37){\includegraphics[scale=0.22]{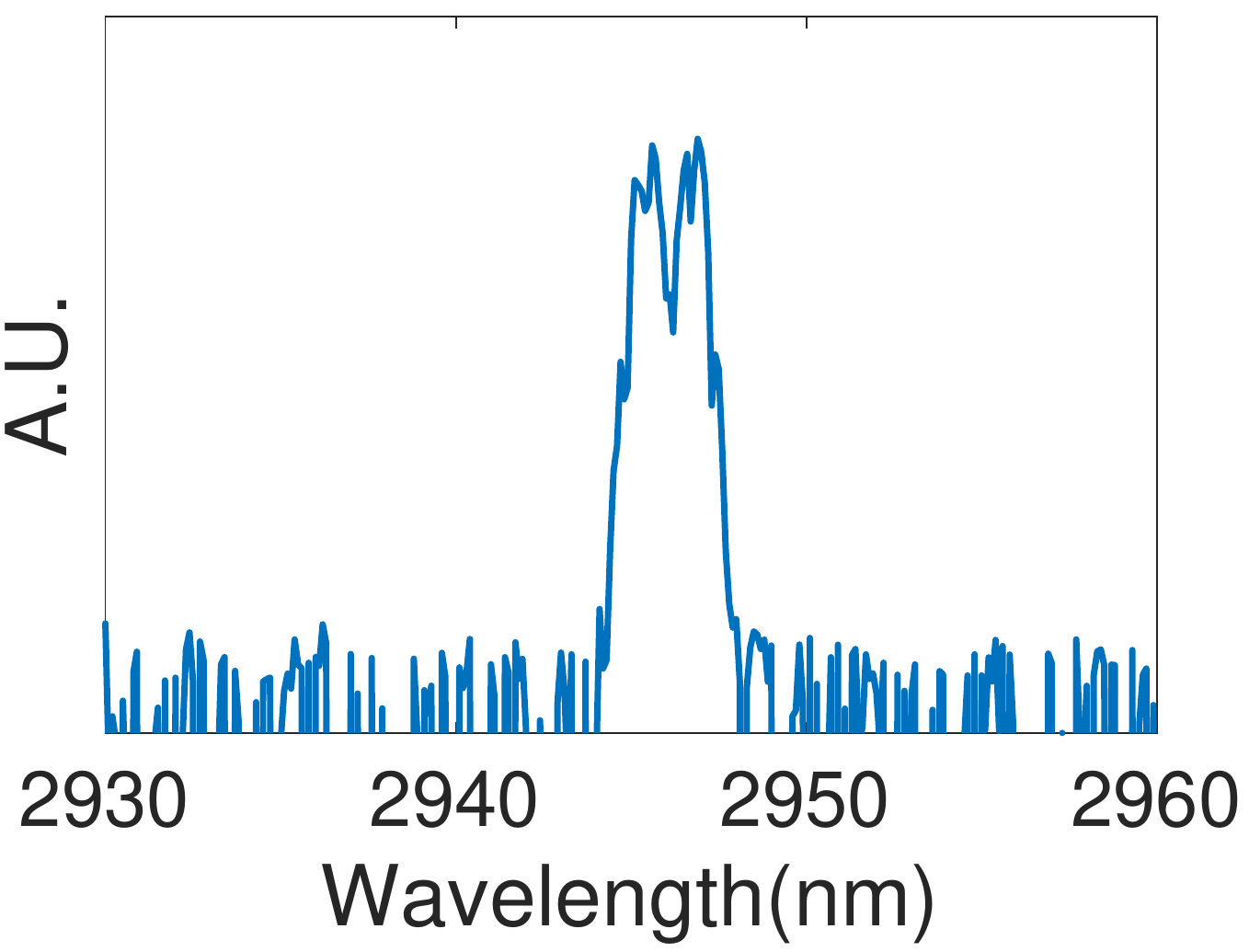}}
	\end{overpic}
	
	\caption{3~\si{\um} laser output power as a function of injected pump power. Slope efficiency '$\eta$' is 14\% with an oscillation threshold of 390~mW. Inset: measured optical spectrum showing narrow linewidth emission centered around 2945~nm}
	\label{fig:slope}
\end{figure}

The laser output power characteristic as a function of injected pump power is presented in Fig.~\ref{fig:slope}. Slope efficiency is 14\% while the injected pump power required for laser oscillation is 390~mW. 
The inset of Fig.~\ref{fig:slope} shows the measured optical spectrum of the laser output, confirming laser emission centered around 2945~nm, as opposed to fluorescence output.
Further power scaling is pump power limited.
While the efficiency achieved is less than that of a recently demonstrated resonantly pumped dysprosium fiber laser~\cite{majewski2016highly}, it is comparable or superior to previous near-infrared pumped systems~\cite{jackson2003continuous,tsang2006efficient}.
The reduction in efficiency from the Stokes limit can largely be attributed to background loss in the fiber, which has been measured to be on the order of $\sim$2~dB/m.
This high loss is due to the early-stage nature of doped InF$_3$ fiber fabrication and is expected to reduce by more than an order of magnitude to comparable values  to doped ZBLAN fiber with ongoing fabrication process improvements.
Of greatest significance however, is that this represents the first demonstration of coherent laser emission from an indium fluoride fiber.
Additionally, demonstration of 3~\si{\um} lasing is a critical step towards realizing a fiber laser beyond 4~\si{\um} should cascade lasing be required to overcome self-termination.

\begin{figure}[!htbp]
	\centering
	\includegraphics[scale=0.5]{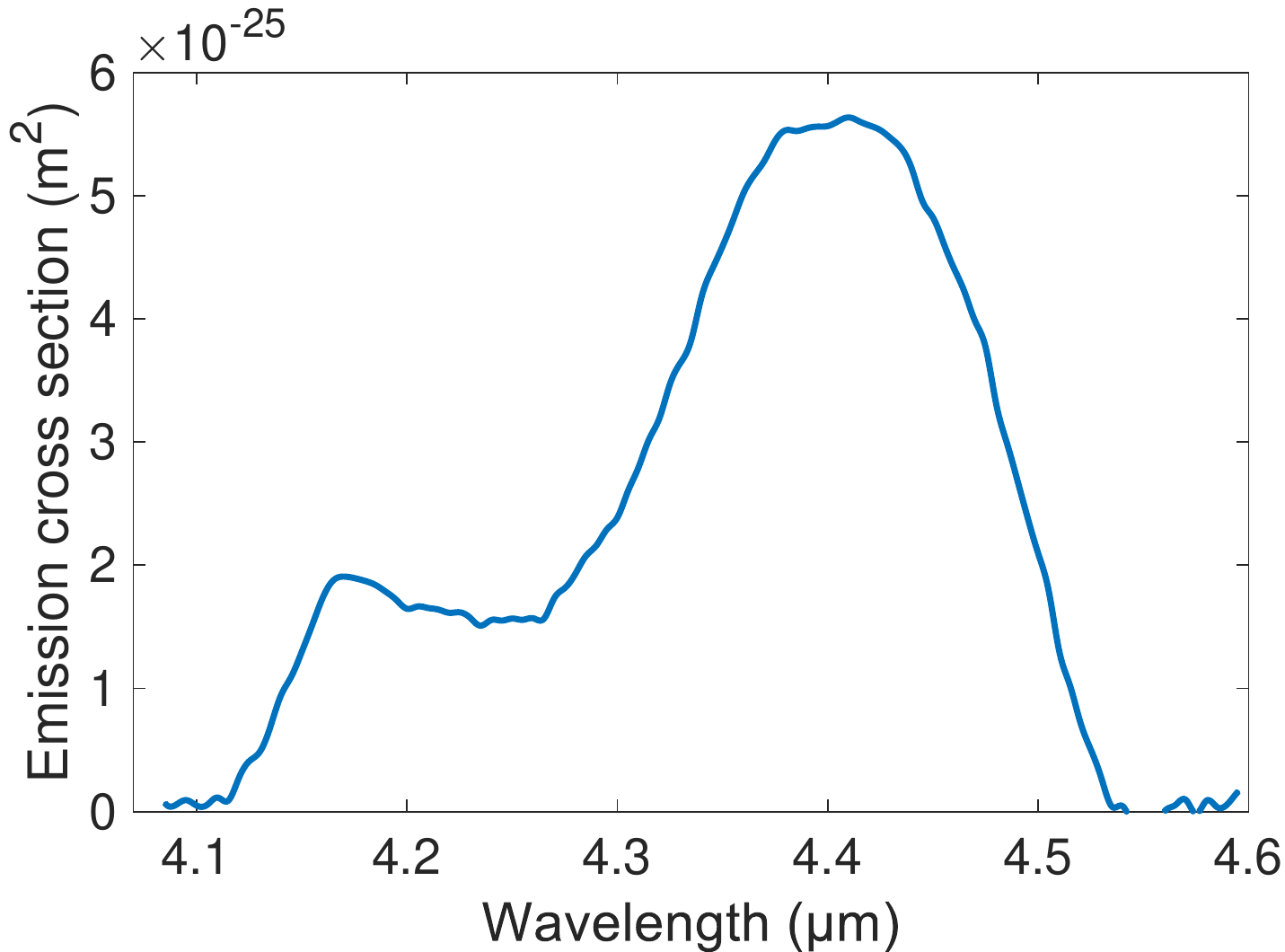}
	\caption{Emission cross section of the $^6H_{11/2}\rightarrow^6H_{13/2}$ transition. Values are calculated from a normalized measured fluorescence spectrum and the F\"uchtbauer--Ladenburg equation}
	\label{fig:4micron}
\end{figure}

To investigate the possibility of emission from the $^6H_{11/2}$ level the cavity mirrors are removed from the arrangement shown in Fig.~\ref{fig:level_setup}(b) as they impart undesirable loss in the 4~\si{\um} region (development of custom mirrors and a suitable resonant cavity for cascade 3 and 4~\si{\um} lasing is a subject of future work).
The Dy:InF$_3$ fiber is again pumped at 1.7~\si{\um} and the collimated output is directed to a grating monochromator.
With an injected pump power of 1~W, emission beyond 4~\si{um} is readily observed at room temperature as seen in Fig.~\ref{fig:4micron}, marking the longest emission wavelength yet observed from a fluoride fiber host.
Measured emission extends from around 4.1 to 4.5~\si{\um}, indicating a similar transition width in this host as seen in dysprosium-doped chalcogenide glass~\cite{tang2012study}.
The emission cross sectional values presented in Fig.~\ref{fig:4micron} are calculated from a normalized fluorescence intensity spectrum $I(\lambda)$, to which the F\"uchtbauer--Ladenburg equation has been applied
\begin{equation}
\sigma_e(\lambda)=\frac{\lambda^4\beta}{8\pi cn^2\tau}\frac{I(\lambda)}{\int I(\lambda)d\lambda}
\end{equation}
where $c$ is the speed of light, $n$ is the refractive index of InF$_3$ ($\sim$~1.5), $\tau$ is the radiative lifetime of the $^6H_{11/2}$ level, and $\beta$ is the branching ratio to the ground state. 
Radiative lifetime and branching ratio values are 13.7~ms and 0.052 respectively, obtained from a Judd-Ofelt analysis in the spectroscopic literature~\cite{gomes2010energy}.
The pronounced dip in the emission spectrum observed around 4.3~\si{\um} can be attributed to strong atmospheric absorption due to carbon dioxide (CO$_2$). 
Interestingly, this feature demonstrates the potential utility of a dysprosium laser source in this region, as it would be able to specifically target this absorption, as well as wholly avoid it if desired.
It should be noted that the fluorescence spectrum is not corrected for this CO$_2$ absorption, thus the true values of emission cross section may be slightly reduced from those presented.  

Raman spectroscopy has shown the maximum phonon energy for indium fluoride glass (509~cm$^{-1}$) is less than that of ZBLAN ($\sim$600~cm$^{-1}$)~\cite{akella1997new}.
The direct observation of 4~\si{\um} fluorescence here we attribute to this difference, as the lower maximum phonon energy translates to increased infrared transparency (see Fig.~\ref{fig:fiber_attn}) as well as a decrease in multiphonon relaxation, which increases the radiative efficiency of the ${^6H_{11/2}\rightarrow^6H_{13/2}}$ transition.
However, there has been some disagreement in the literature here, as Quimby and Saad reported measurements of the $^6H_{13/2}$ excited state lifetime in indium and zirconium fluoride hosts, concluding that the non-radiative decay rates for the two glasses are nearly identical~\cite{quimby2017anomalous}.
These measurements were conducted in bulk doped glass, and the specific InF$_3$ composition is not given; therefore as a point of direct comparison, we experimentally measure the $^6H_{13/2}$ lifetime in both the InF$_3$ fiber here, and the ZBLAN fiber used in previous work~\cite{Majewski:18}, which is identical in doping concentration and fiber geometry.

Excited state lifetime measurements in the time domain often require a high degree of excitation in the form of short intense pump pulses.
Additionally, high sensitivity and high speed detection of fluorescence is necessary.
Alternatively, fluorescence lifetime can be measured in the frequency domain, with greatly reduced excitation and sensitivity requirements~\cite{brunel1996simple}.
This method involves modulating the pump beam and characterizing the phase of the fluorescence emission as a function of modulation frequency.

\begin{figure}[!htbp]
	\centering
	\begin{overpic}[trim= 0 -5mm 0 -5cm, scale=0.5]{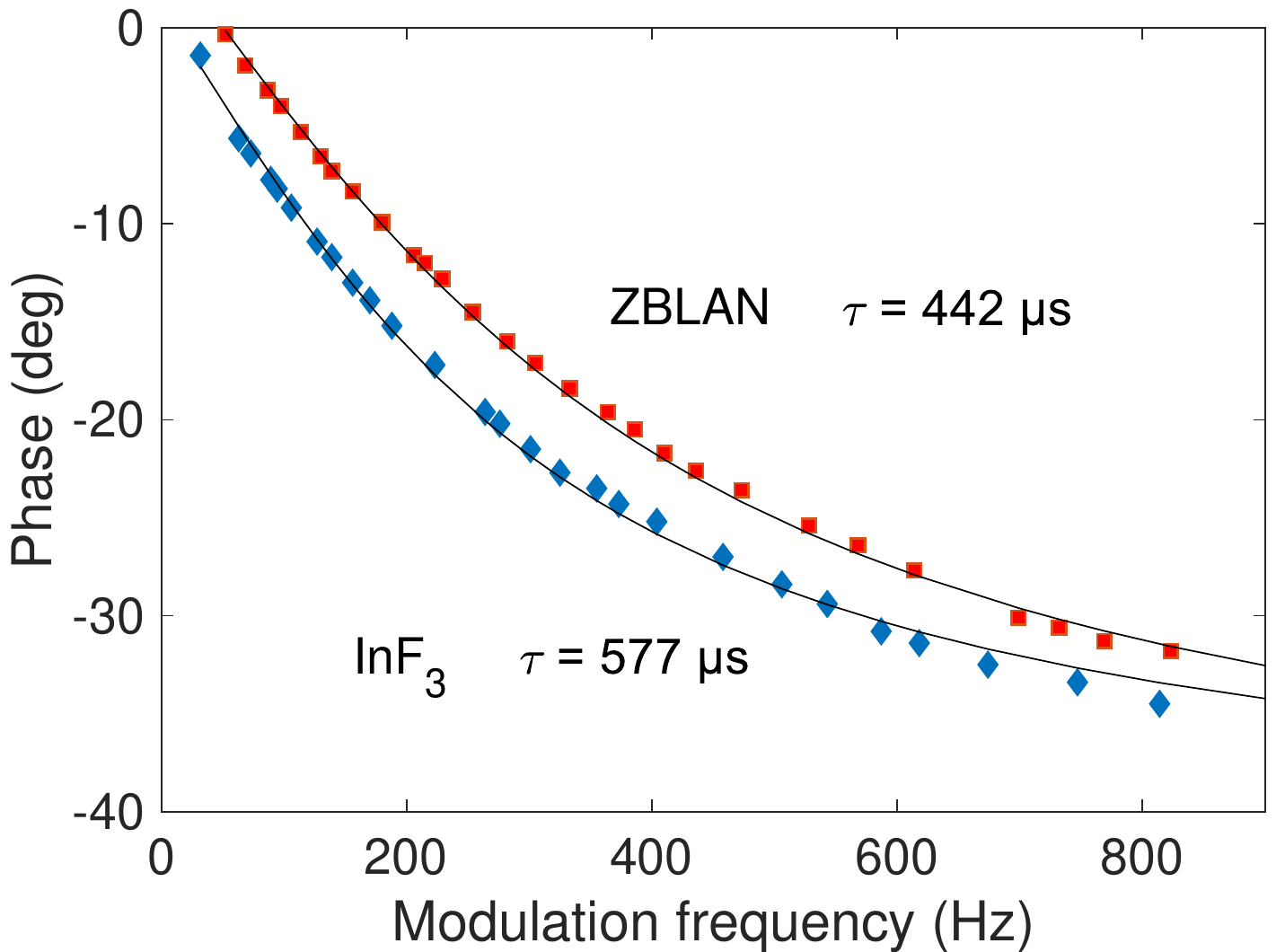}
		
		\put(-5,75){\includegraphics[scale=0.2]{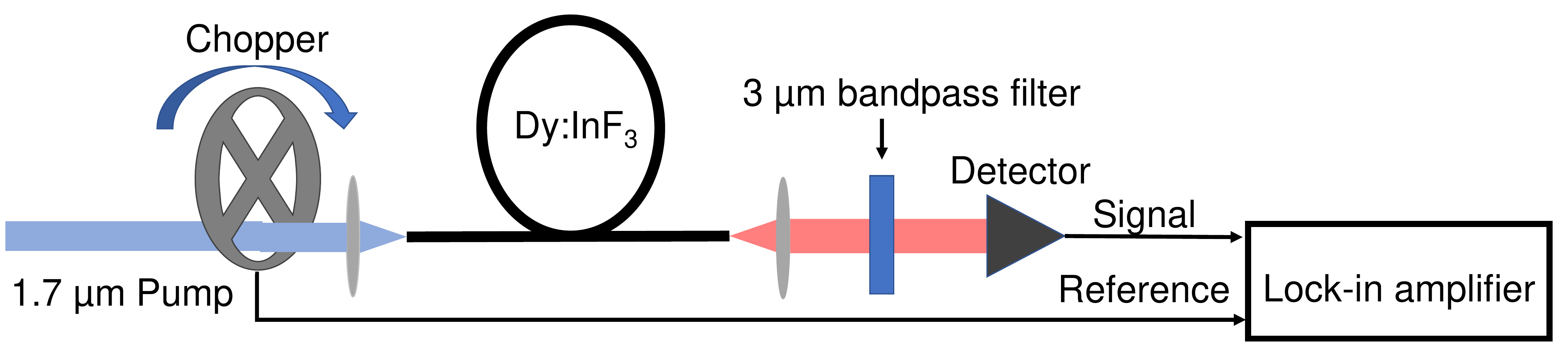}}
		\put(50,70){\textbf{(a)}}
		\put(50,-2){\textbf{(b)}}
	\end{overpic}
	
	\caption{(a) Setup for the frequency domain measurement of $^6H_{13/2}$ lifetime. (b) Measured phase response of the 3~\si{\um} fluorescence. Least squares fits to $\phi(\omega)=\arctan(\omega\tau)$ yield total lifetimes of 442 and 577~\si{\us} for zirconium and indium fluoride respectively.}
	\label{fig:lifetime}
\end{figure}
The laboratory setup implemented for this measurement is shown in Fig.~\ref{fig:lifetime}(a). 
A mechanical optical chopper modulates the pump beam and provides the reference frequency for a lock-in amplifier.
Output from the fiber is first bandpass filtered to reject unabsorbed pump light, and then measured with a PbSe photoconductive detector connected to a lock-in amplifier. 
The phase response $\phi$, of the photodetector signal is then measured and fit to the functional form
\begin{equation}
\phi(\omega)=\arctan(\omega\tau)
\end{equation}

where $\omega$ is the modulation frequency (in radians/s), and $\tau$ is the total lifetime (derivation in Ref.~\cite{brunel1996simple}).
The decision to measure the 3~\si{\um} lifetime is a direct result of this method requiring modulation frequency of at least $1/\tau$.
From measurements in Dy:ZBLAN~\cite{gomes2010energy}, the 4~\si{\um} lifetime is expected to be on the \si{\us} scale, requiring modulation of up to 1~MHz, thus mechanical chopping is not suitable.

The measurement results are presented in Fig.~\ref{fig:lifetime}(b), and show good agreement to the expected functional form.
The lifetime estimate from these fits show a clear difference between the two host glasses, with indium fluoride exhibiting a substantially longer fluorescence lifetime of 577~\si{\us} compared to 442~\si{\us} in ZBLAN.
These results are in-line with the expectation that lower maximum phonon energy of the host glass should translate to reduced non-radiative decay rates and increased fluorescence lifetime.
One consequence of this has already been shown in the room-temperature observation of fluorescence beyond 4~\si{\um}, but the increase in lifetimes may prove InF$_3$ to be a superior host to ZBLAN for 3~\si{\um} fiber lasers as well.

In conclusion we have demonstrated MIR lasing from an indium fluoride fiber for the first time.
Coherent emission at 2.95~\si{\um} from the $^6H_{13/2}\rightarrow^6H_{15/2}$ transition of dysprosium is achieved by near-infrared pumping at 1.7~\si{\um}, resulting in conversion efficiency of 14\% and a pump-power limited output.
Direct measurements of the excited state lifetime using a frequency domain method reveal a substantial increase in lifetime in InF$_3$ over ZBLAN fiber.
This implies a reduced influence of multiphonon effects, which is further evidenced by the direct observation of fluorescence from the $^6H_{11/2}\rightarrow^6H_{13/2}$ transition that is non-radiatively quenched in ZBLAN.
Emission spans nominally from 4.1 to 4.5~\si{\um} and represents the longest emission wavelength yet achieved from a fluoride-based host glass.
These results show promise for the realization of fiber lasers beyond the historical 4~\si{\um} wavelength barrier.

\section*{Funding Information}
Australian Research Council (ARC) (DP140101336, DP170100531).
RIW acknowledges support through an MQ Research Fellowship.
\bigskip


\end{document}